\documentclass[pre,twocolumn,superscriptaddress]{revtex4}

\usepackage{epsfig}
\usepackage{latexsym}
\usepackage{amsmath}
\usepackage{amssymb}
\usepackage{amsfonts}
\usepackage{wrapfig}
\usepackage[left]{lineno}
\usepackage{url}
\usepackage{graphicx}
\usepackage{dcolumn}
\usepackage{bm}
\usepackage{psfrag}
\usepackage{physics}
\usepackage{hyperref}
\usepackage{soul} 
\usepackage{lipsum}

\usepackage[normalem]{ulem}  
\usepackage[dvipsnames]{xcolor} 

\definecolor{darkgreen}{rgb}{0.0, 0.4, 0.0}
\definecolor{Mycolor1}{rgb}{0.6,0.,0.7}
\newcommand{\old}[1]{}

\renewcommand{\k}{{\rm k}}
\renewcommand{\c}{{\rm c}}

\begin{document}


\title{Shape matters: Competing mechanisms of particle 
shape segregation}

\author{D. Hern\'{a}ndez-Delfin} 
\affiliation{Departamento de F\'{\i}sica y Matem\'{a}tica Aplicada, 
Universidad de Navarra, P.O. Box. 177, E-31080  Navarra, Spain}
\affiliation{BCAM - Basque Center for Applied Mathematics,
Mazarredo, 14 E48009 Bilbao, Basque Country – Spain.}

\author{D.R. Tunuguntla}
\affiliation{Multiscale Mechanics, Department of Thermal and Fluid Engineering, Faculty of Engineering Technology, MESA+, University of Twente, P.O. Box. 217, 7500 AE Enschede, The Netherlands}
\author{T. Weinhart} 
\affiliation{Multiscale Mechanics, Department of Thermal and Fluid Engineering, Faculty of Engineering Technology, MESA+, University of Twente, P.O. Box. 217, 7500 AE Enschede, The Netherlands}
\author{R.C. Hidalgo}
\email{raulcruz@unav.es}
\affiliation{Departamento de F\'{\i}sica y Matem\'{a}tica Aplicada, 
Universidad de Navarra, P.O. Box. 177, E-31080  Navarra, Spain}
\author{A.R. Thornton}
\email{a.r.thornton@utwente.nl}
\affiliation{Multiscale Mechanics, Department of Thermal and Fluid Engineering, Faculty of Engineering Technology, MESA+, University of Twente, P.O. Box. 217, 7500 AE Enschede, The Netherlands}

\date{\today}

\begin{abstract}
It is well-known that granular mixtures that differ in size or shape segregate when sheared. In the past, two mechanisms have been proposed to describe this effect, and it is unclear if both exist. To settle this question, we consider a bidisperse mixture of spheroids of equal volume in a rotating drum, where the two mechanisms are predicted to act in opposite directions. 
We present the first evidence that there are two \emph{distinct} segregation mechanisms driven by relative \emph{over-stress}. Additionally, we showed that for non-spherical particles, these two mechanisms can act in different directions leading to a competition between the effects of the two. As a result, the segregation intensity varies non-monotonically as a function of $AR$, and at specific points, the segregation direction changes for both prolate and oblate spheroids, explaining the surprising segregation reversal 
 previously reported. 
Consistent with previous results, we found that the kinetic mechanism is dominant for (almost) spherical particles.  Furthermore, for moderate aspect ratios, the kinetic mechanism is responsible for the spherical particles segregation to the periphery of the drum, and the gravity mechanism plays only a minor role. Whereas, at the extreme values of $AR$, the gravity mechanism notably increases and overtakes its kinetic counterpart.


\end{abstract}
\maketitle

\section{Introduction}
Anyone who has shaken a box of cereal has observed segregation/de-mixing. This is the so-called Brazil-nut problem, which still has many open questions despite being heavily studied \cite{RosatoStrandburgPrinzandSwenden1987,Gajjar:2021wq}. In general, segregation occurs due to differences in physical properties, ranging from shape and size to the coefficient of restitution and sliding friction. However, in sheared dense granular flows, it is the difference in size that primarily drives segregation \cite{Gray2018}.


Size-based segregation in sheared dense granular flows has recently attracted a lot of scientific attention, utilizing a variety of different approaches and geometries.  For example, van der Vaart {\it et al.} \cite{Vaart2015} experimentally studied  particle size segregation in a shear box, looking at individual small and large particle dynamics in a system under oscillatory shear. In contrast, Jing {\it et al.} \cite{Jing2017} used particle simulations in a periodic chute to show that individual large particles carry higher contact forces. This {\it over-stress} (or \emph{over-pressure}) acts as a mechanism that drives these large particles upwards.


The idea of large particle {\it over-stresses} measured by Jing {\it et al.} \cite{Jing2017} (and many others) had previously been theoretically postulated by Gray \& Thornton \cite{Gray2005theory,ThorntonGrayHogg2006}. However, they do not express the {\it over-stress} in terms of particle-size, which left open two questions: How does the over-stress scale with particle size-ratio; and, how do you define size? 
In the original model, the over-stress is assumed to be proportional to the hydrostatic load (pressure), which for most geometries scales with gravity. However, in dynamic cases, Fan \& Hill \cite{fan2011phase,Fan2011} showed a second source of over-stress called kinetic stress, which is due to difference in velocity fluctuations and does not scale with gravity.

For polydisperse mixtures of spheres, Tunuguntla et al.~\cite{Tunuguntla2016} developed a novel micro-macro analysis technique, based on coarse-graining (CG)\,\cite{Goldhirsch2010,Weinhart2012b,Weinhart2013,Richard2015,Artoni2019}. 
{This is required as it consistently splits the stress of a contact between different sizes of particles, which is not guaranteed with simpler methods, e.g., binning.} CG is used to obtain macroscopic properties of granular flows from microscopic details. This technique is widely used as a micro-macro mapping procedure presenting advantages compared to other, simpler methods, namely binning and the method of planes \cite{Tunuguntla2017}. Using the technique, Tunuguntla et al. \cite{Tunuguntla2017} showed that, in bidisperse mixtures flowing down inclined planes, the over-stress in the kinetic stress is far greater than the contact stress. However, both act in the same direction. Hence it is very hard to distinguish the effects and raises the questions: Is there one or two segregation mechanisms, and what happens for non-spherical particles?

Lu and Müller \cite{Lu2020} investigated mixtures of spherical and non-spherical particles in a rotating drum. They explored mixtures with different blockiness; always, non-spherical particles segregated towards the center. Various other authors have also studied segregation in rotating drums with shapes such as cuboids and spheres \cite{pereira2017segregation} and rods \cite{zhao2018simulation, Jones2020}. In these studies, the particles differed in volume and mass. Recently, He {\it et al.} \cite{He2019} numerically studied a bidisperse mixture of ellipsoids in a rotating drum, keeping the particle volume constant. Under these conditions, segregation is exclusively a result of the differences in shape. Astonishingly, they showed that the segregation direction changes depending on the shape differences. They attributed the results to two competing segregation mechanisms. The first one is related to the fact that spheres have better flowability than non-spherical particles, making spheres flow along the strongly sheared drum periphery while ellipsoids deposited at the center as they dissipate more energy in the flowing layer. They described a second mechanism by observing that particles tend to orient more against the flow direction as they become more non-spherical, allowing spheres to percolate towards the core as ellipsoids offer less resistance.


\begin{figure}
\begin{center}
\includegraphics[scale=1]{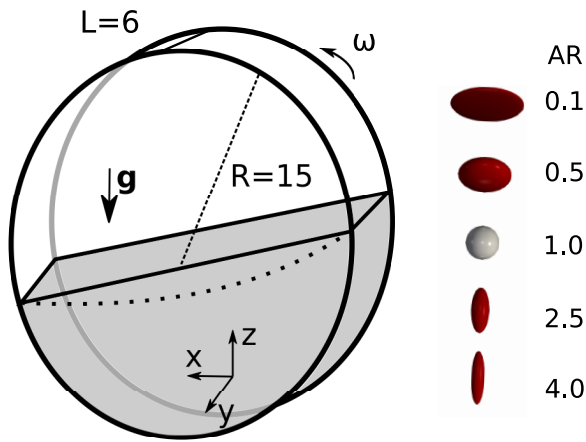}
\caption{Sketch of the numerical system; the dashed line represents the boundary between the flowing layer and the solid body. 
In addition, the inset shows a representation of the particle shapes ($AR \in \lbrace0.1,0.5,1.0,2.5,4.0\rbrace$).}
\label{fig:newFig_skecth}
\end{center}
\end{figure}

In this work, inspired by the segregation reversal reported by He {\it et al.}  \cite{He2019} for equal-volume particles and equipped with the state-of-art particle analysis tool MercuryCG \cite{Tunuguntla2016}, we thoroughly analyse the segregation behaviour of spheroids for different aspect ratios within the existing theoretical frameworks. Remarkably,  we 
demonstrate 1) the segregation reversal is predicted by the idea of an \textit{over-stress}, and 2) 
there are two distinct stress segregation mechanisms, which even compete with each other.

\section{Numerical model (DEM)}

The model consists in a non-dimensional system involving a bidisperse mixture of spherical and ellipsoidal particles in a rotating drum, subjected to a gravitational field of magnitude $g = 1$. Figure \ref{fig:newFig_skecth} illustrates a sketch of the numerical setup, a drum is of radius $R = 15d$ and depth $L = 6d$, with periodic boundary conditions in the transverse direction ($y$-{axis}). Gravity acts in $z$-direction. 
The main goal of the simulations is to study segregation as a result of differences in shape only, the material densities of both constituents are equal, and spheroid semi-diameters $a$, $b$, and $c$ are chosen in order to keep equal volumes, and therefore equal masses. 
Thus, each sphere ($a = b = c$) has diameter $d =1$ and mass $m = 1$, respectively. 
Similar to He {\it et al.} \cite{He2019}, we define an aspect ratio $AR=a/c$ and set $a = AR^{2/3}/2$ and $b = c = AR^{-1/3}/2$, such that the spheroids have the same volume as the spheres.
If $AR < 1$, the mixture is made of spheres and prolate spheroids; if $AR = 1$, both species are spheres; otherwise, the bed is made of spheres and oblate spheroids. The studied range of $AR$ was from $0.1$ to $4.0$ at $0.1$ intervals, with a total number of $N = 2000$ particles, half of which belong to each one of the constituents. 

The drum walls are constructed by placing spherical particles of diameter $d/2$ onto a cylindrical surface in a $L \times 2\pi R$ grid and forcing them to rotate at a given angular velocity around the $y$-{\it axis}. Constructing the drum from particles creates a rough surface with sufficient friction to avoid slipping between the drum and the bulk particles. The rotation velocity $\omega$ is set such that the non-dimensional Froude number $F=\omega^2 R/g$, which characterizes the ratio of centrifugal to gravitational forces, is $F = 0.01$. This value is close to the upper limit of the rolling regime, which is one of the six categories of rotating drum flows and one of the most relevant for industrial applications \cite{Govender2016}. Under this regime, the granular flow involves two distinct zones, a thin flowing layer at the surface and a passive solid body rotation below.

Simulations were carried out in \href{http://www.mercurydpm.org/}{MercuryDPM} \cite{MercuryPaper2020}, an open-source code for discrete element modeling (DEM) particle simulations. To model forces among particles, the linear spring-dashpot contact model is used, with a restitution coefficient $e_r = 0.1$, collision time $t_c = 0.05$, and Coulomb friction coefficient $\mu = 0.5$. In order to find the contact point and compute the overlap between two particles, we consider spheroids and spheres as a subset of superquadric particles. The implementation is similar to \cite{Podlozhnyuk2017}, describing the surface of superquadrics 
with $n_1 = n_2 = 2$ for ellipsoidal shapes; inset in Figure \ref{fig:newFig_skecth} illustrates particle shapes for $AR \in \lbrace0.1,0.5,1.0,2.5,4.0\rbrace$.
As a starting point, the $N$ particles are randomly placed inside the drum, and we wait until the system is relaxed before starting to rotate the drum. Setting the time step as $\Delta t = t_c/50$, we analyze the results after ten rotations, at which time He {\it et al.} \cite{He2019} observed appreciable differences regarding the components  segregation. 

\section{Obtaining continuum fields}

The previously mentioned continuum models are all expressed in terms of Eulerian continuum properties like stress and volume fraction. However,  the DEM algorithm resolves the discrete Lagrangian trajectories of all the simulated particles,  providing their location and contact network  with a given time resolution. We employ a coarse-graining methodology to post-process this discrete data to the required continuous fields which are required for the theoretical models \cite{Goldhirsch2010,Weinhart2013,Richard2015,Artoni2019}. Analogous to \cite{Tunuguntla2017}, the simulated system includes three different types of constituents (bulk spheres) type-$s$, (bulk spheroids) type-$e$, and (boundary) type-$b$. The interstitial pore space is considered as a zero-density passive fluid. Following Tununguntla {\it et al.} \cite{Tunuguntla2017} we define the subsets of spherical particles, $\mathcal{F}^s$, elliptical particles,  $\mathcal{F}^e$, boundary particles,  $\mathcal{F}^b$, and their union  $\mathcal{F} = \mathcal{F}^s \cup \mathcal{F}^e \cup \mathcal{F}^b$, where each particle $i \in \mathcal{F}^\nu$ has a mass $m_i$, a center of mass $\vb{r}_i$ and velocity $\vb{v}_i$.

According to~\cite{Tunuguntla2017,Goldhirsch2010,Weinhart2013,Richard2015,Artoni2019}, the mass density corresponding to type-$\nu$ constituent, $\rho^{\nu}(\vb{r},t)$, at $\vb{r}$ and time $t$ is defined by
\begin{equation}
\rho^{\nu}\left(\vb{r},t\right) = \sum_{i \in \mathcal{F}^{\nu}} m_i \psi\left(\vb{r}-\vb{r}_i(t)\right)
\end{equation}
\noindent and $\psi\left(\vb{r}-\vb{r}_i(t)\right)$, henceforth $\psi_i$ for simplicity, is an integrable coarse-graining function. In the present work, we choose the Lucy-polynomials with a width of 1, i.e., the diameter of the spherical particles, as recommended by \citet{Tunuguntla2016}.

To satisfy the mass and momentum conservation equations, expressions for the other type-$\nu$ partial quantities are

\begin{eqnarray}
\begin{aligned}
\vb{u}^{\nu} &= \frac{1}{\rho^{\nu}}\sum_{i \in\mathcal{F}^{\nu}} m_i \vb{v}_{i} \psi_i,\\
\pmb{\sigma}^{k, \nu} &= \sum_{i \in\mathcal{F}^{\nu}} m_i \vb{v}_{i  }' \vb{v}'_{i  } \psi_i, \\
\quad \pmb{\sigma}^{c, \nu} &= \sum_{i \in \mathcal{F}^{\nu}} \sum_{j \in \mathcal{F}} \vb{f}_{ij  } \vb{b}_{ij  } \psi_{ij},
\end{aligned}
\end{eqnarray}
where $\vb{v}'_i$ is the fluctuation of the velocity of particle $i$ with respect to the mean field, $\vb{v}_{i}'  =\vb{u} - \vb{v}_{i}$, $\vb{f}_{ij}$ and $\vb{b}_{ij}$ denote the force between particle $i$ and $j$ and the  branch vector between particle $i$ and its contact point with particle $j$, respectively. Finally, $\psi_{ij}$ is the lineal integral along $\vb{b}_{ij}$, $\psi_{ij} = \int_0^1  \psi(\vb{r} - \vb{r}_i + s \vb{b}_{ij} ) ds$, which distributes the contact stress proportionally, subjected to the corresponding branch fraction of each constituent.

Following the previous procedure, all relevant macroscopic fields were extracted from the DEM data. The fields were averaged over the depth of the drum and in time for a period of 1.5 rotations, starting from the tenth rotation, with a temporal resolution $\Delta t_{CG} = 0.01 \sqrt{d/g}$. 

\section{Results and discussion}

\begin{figure}
\begin{center}
\includegraphics[scale=.5]{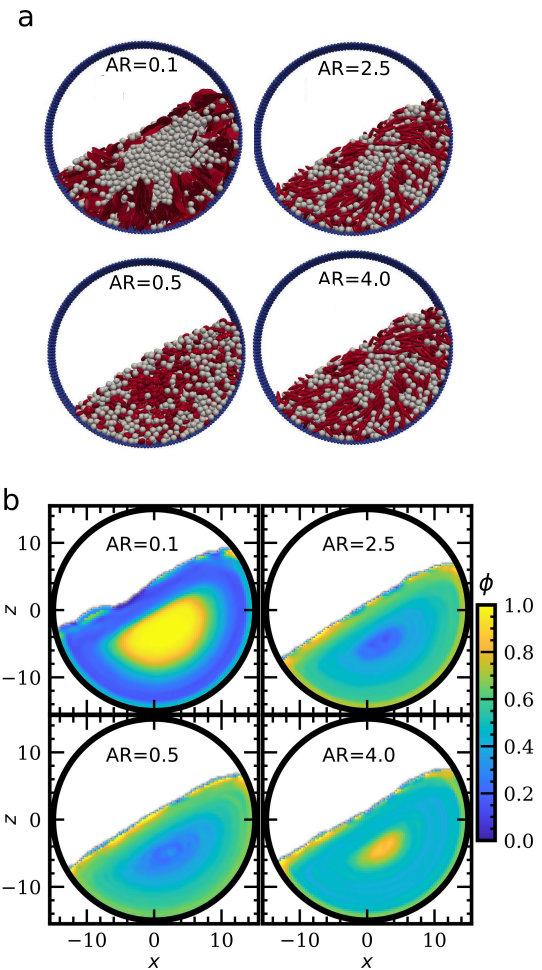}
\caption{(a)  
Snapshots of the mixtures $AR = \lbrace$0.1,0.5,2.5,4.0$\rbrace$ after ten rotations ($t \approx$ 2433). The spherical particles are colored gray, while elongated particles are red.
(b) The concentration of spheres $\phi_s(x,z)$ obtained from density CG-fields for the same $AR$ values of (a).}
\label{fig:newFig1_cg}
\end{center}
\end{figure}

We performed DEM simulations of a rotating drum by exploring the impact of the particle aspect ratio $AR$ on the segregation process. To this end, a simple visual inspection of the system can tell whether the particles tend to group by shapes after a while. Figure \ref{fig:newFig1_cg} a) depicts the mixture states after ten rotations for $AR = \lbrace$0.1,0.5,2.5,4.0$\rbrace$. As can be clearly seen, segregation is evidenced in all these cases at different degrees. The most extreme mixture ($AR =$ 0.1) shows that most of the spherical particles are in the core of the drum, while non-spherical ones are in the periphery. In case $AR =$ 0.5, the segregation inverts despite being less intense. The inversion of the segregation also manifests in mixtures that contain prolate ellipsoids. 
Similar segregation reverse was previously found, e.g., the called Reverse Brazilian Effect (RBE) \cite{Hong2001,Breu2003} or the recent result in reference \cite{Duan2021}, where the inversion is obtained by varying size and density simultaneously. Here, the most striking about this result is that the segregation reverse occurs only with varying the particle shape (see also Ref.\cite{He2019}). 

{

Here we will define two types of volume fraction the normal volume fraction per unit $\varphi^\nu$ mixture volume, which is simply the local ratio of volume of constituent $i$, over total volume $\varphi^\nu=V^{i}/V$.   From this we can define volume fraction per unit granular volume that is $\phi^\nu=\varphi^\nu/(\varphi^s+\varphi^e)$, which has the property $\phi^e + \phi^s = 1$.}
Figure \ref{fig:newFig1_cg} b) shows as color-map the  volume fraction per solid volume of spheres $\phi^s(x,z)$. 
{It resembles the spatial distribution of spheres and ellipsoids, depicted in Figure \ref{fig:newFig1_cg} a).}
One can notice that $\phi^s$ is non-homogeneous, with higher values in the core of the drum for mixtures with extremely elongated spheroids ($AR =$ 0.1 and $AR =$ 4.0). In contrast, the opposite behavior is observed for $AR$ values close to 1, where the mixture is homogeneous. 

Steeping forward, using the quantities $\phi^\nu$ and the bulk density $\rho(\vb{r})$, we adapt the method of Arntz et al. \cite{Arntz2008,Arntz2014} to quantify segregation accurately. From statistical mechanics, the mixing entropy is defined as
\begin{equation}
    \tilde{M} = \sum_{\nu\in\{s,e\}} \int_{\mathbb{R}^3} \rho(\vb{r}) \phi^{\nu}(\vb{r})\ln{\phi^{\nu}(\vb{r})} \dd\vb{r}.
\end{equation}
Next, we define a segregation index $S$ that measures not only the segregation intensity but also the direction in which segregation occurs in a rotating drum (inwards or outwards),
\begin{equation}
S  = \left\{
        \begin{array}{ll}
            M - 1& \text{if spheres segregate inwards}, \\
            1 - M & \text{if spheres segregate outwards}.
        \end{array}
    \right.
\label{eq:segregation}
\end{equation}
Thus, a positive value of $S$ indicates that spheres tend to segregate away from the core.

\begin{figure}
\begin{center}
\includegraphics[scale=1]{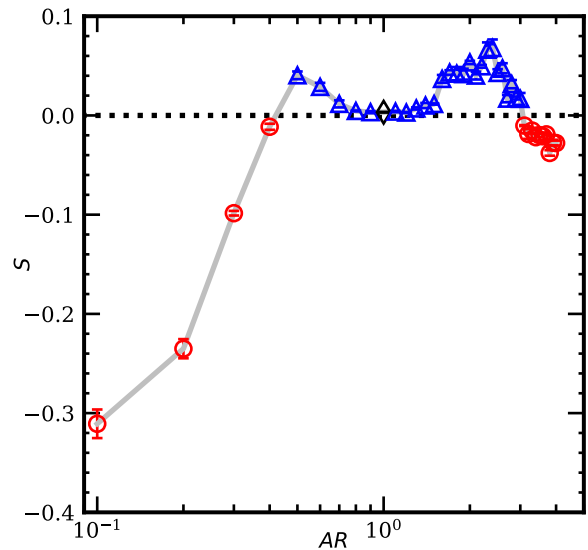}
\caption{The segregation index $S$ as a function of the aspect ratio $AR$, which has been computed for each of the three half rotations and averaged. The red circles illustrate the aspect ratios where spheres segregate to the center of the drum, whereas the blue triangles indicate that spheres go to the periphery. For $AR = $ 1.0, a particular marker is assigned (black diamond). In all cases, error bars account for the standard deviation of the mean value.}
\label{fig:segregation}
\end{center}
\end{figure}

\begin{figure}
\begin{center}
\includegraphics[scale=1]{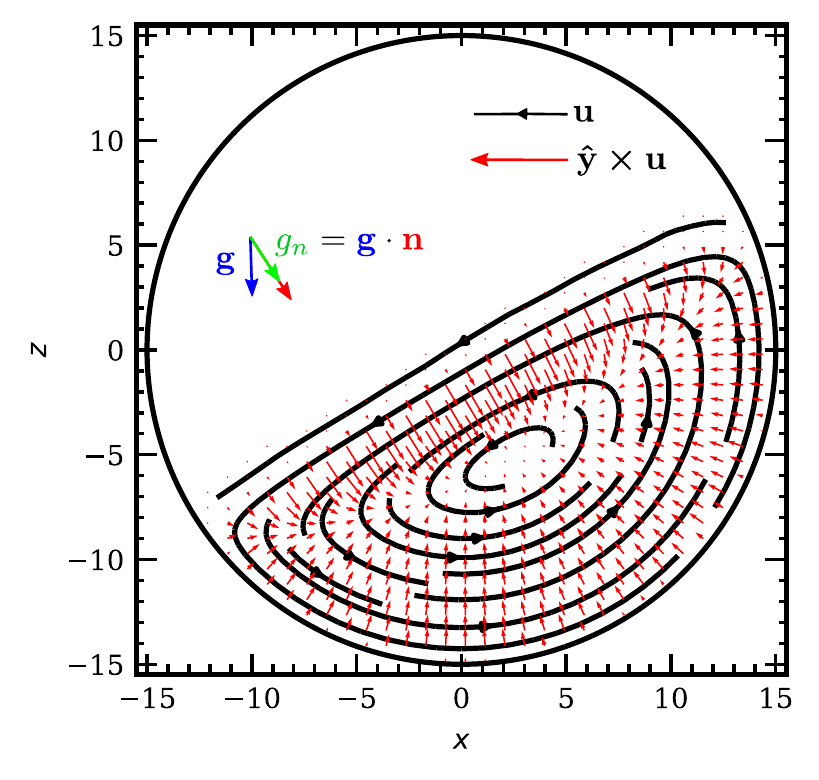}
\caption{Drawing representing the velocity field $\vb{u}$ (black streamlines) and its perpendicular field $\hat{\vb{y}} \cross \vb{u}$ (red arrows). Complementary, gravity $\vb{g}$ and its projection along a vector $\vb{n} = \hat{\vb{y}} \times \vb{u}/{|\vb{u}|}$ are represented, $g_{\rm n}=\vb{g}\cdot\vb{n}$.}
\label{fig:stream}
\end{center}
\end{figure}

Going further, we show the spatially-averaged segregation index Eq.~\ref{eq:segregation} as a function of $AR$ in figure~\ref{fig:segregation}. 
For $AR$ values close to 1.0, the values of $S$ are positive (labelled with blue triangles), indicating that the spheres tend to segregate to the drum periphery, whereas $S$ is negative for extreme $AR$ values (labelled with red circles); thus, the spheres segregate towards the core.

For the case of prolate spheroids, $S$ reaches a maximum local value at $AR \approx$  2.5 and flips direction at $AR \approx$ 3.1, suggesting that a new effect starts to play a role in the behavior of the system. Similarly, for oblate spheroids, $S$ reaches a maximum local value at $AR \approx $ 0.5 and flips direction at $AR \approx$ 0.4. At $AR =$ 1 (labeled with a black diamond), $S \approx$ 0, a result expected beforehand because segregation must not occur for similar constituents. These results are similar to \cite{He2019} and indicate that the segregation changes direction at specific $AR$ values. 

\subsection{Continuum analysis of particle segregation} 

\begin{figure*}[ht]
\begin{center}
\includegraphics[scale=0.9]{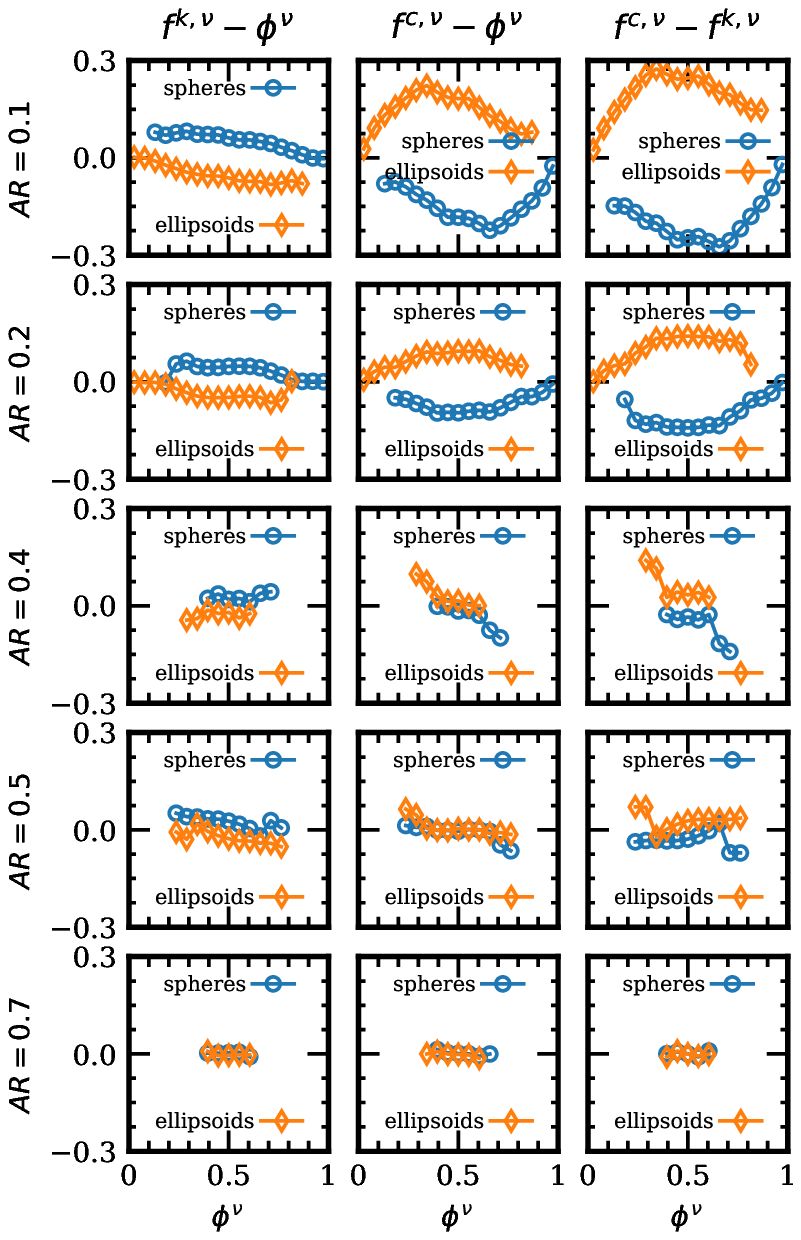}\includegraphics[scale=0.9]{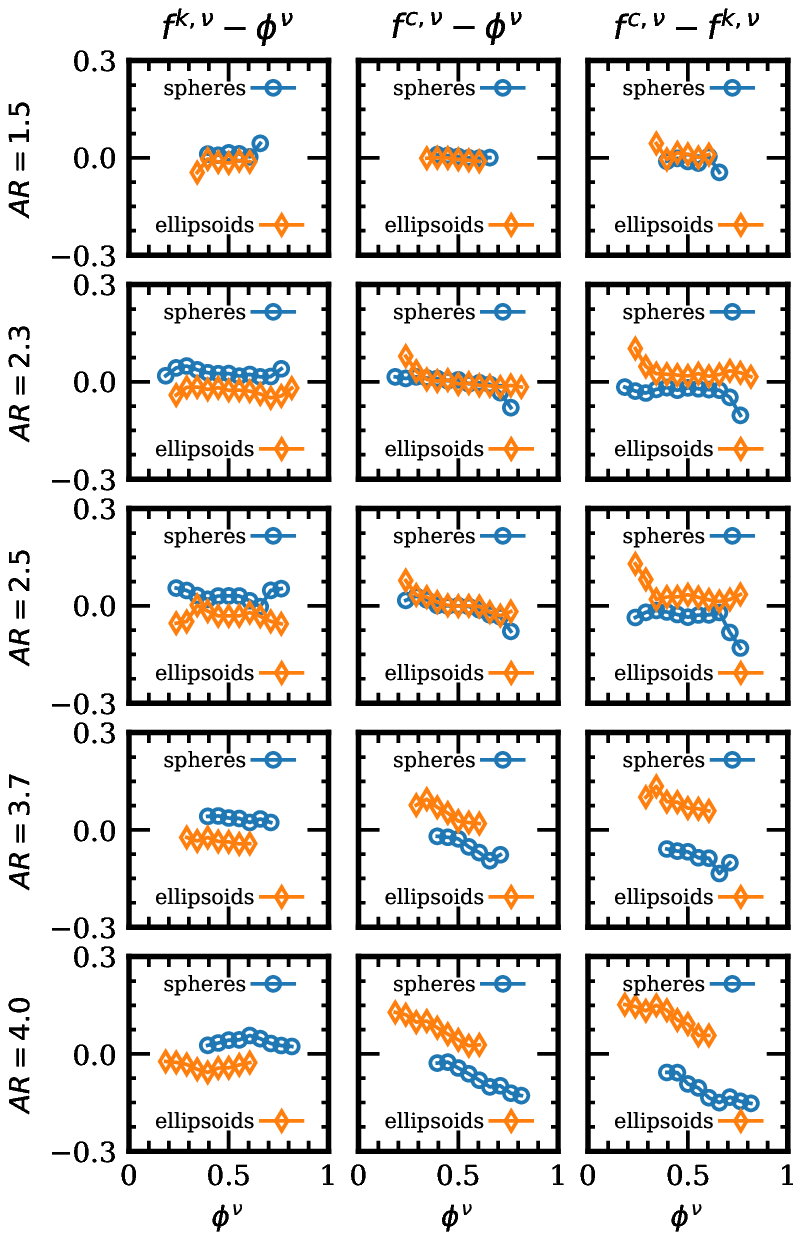}
\caption{Every row illustrates the profiles of $(f^{\k,\nu} - \phi^{\nu})$ (left column), $(f^{\c,\nu} - \phi^{\nu})$ (center column), and $(f^{\c,\nu} - f^{\k,\nu})$ (right column) as a function of $\phi^{\nu}$ for oblate $AR \in \lbrace0.1,\, 0.2,\, 0.4,\, 0.5,\, 0.7\rbrace$ and prolate $AR \in \lbrace1.5,\, 2.3,\, 2.5,\, 3.7,\, 4.0\rbrace$ ellipsoids, respectively.
} 
\label{fig:prefactors_oblate}
\end{center}
\end{figure*}

Next, we quantify the segregation mechanisms and clarify the origin of the change in the segregation direction, employing the theoretical framework introduced by Fan and Hill \cite{Fan2011}.
This formulation is based on the partial momentum and mass conservation equations for individual constituents of a mixture \cite{morland1992flow},
\begin{subequations}
\begin{align}\partial _{t}\rho ^{\nu }+\nabla \cdot \left( \rho ^{\nu }\vb{u}^{\nu }\right) &=0,\\
\rho ^{\nu }\left( \partial _{t}\vb{u}^{\nu }+\vb{u}^{\nu}\cdot \nabla \vb{u}^{\nu }\right) &=-\nabla \cdot \pmb{\sigma} ^{\nu }+\rho ^{\nu }\vb{g}+\pmb{\beta} ^{\nu },\label{eq:momentum}
\end{align}\label{eq:continuity}
\end{subequations}
where $\rho ^{\nu}$, $\vb{u}^{\nu }$, and $\pmb{\sigma}^{\nu}$ are the density, velocity, and stress of the constituent $\nu$, respectively. $\vb{g}$ is the gravity vector, and $\pmb{\beta}^{\nu}$ is the inter-constituent drag force that obeys $\sum_{\nu}\pmb{\beta}^{\nu} = \pmb{0}$. 
{We use the coarse-graining formulae from section III in order to directly obtain these mixture variables from the discrete particle data.}

We consider systems in which the velocities and partial densities approach steady state long before the segregation profile equilibrates, such that the temporal derivatives in \eqref{eq:momentum} become negligible. Moreover, we assume that segregation occurs in the direction of the shear plane orthogonal to the barycentric velocity, $\vb{u} = \sum_{\nu} \rho^{\nu} \vb{u}^{\nu}/\rho$. 

Thus, if the drum is rotating around the $y$-axis, then the direction of segregation is $\vb{n} = \vb{\hat{y}} \times \vb{u}/{|\vb{u}|}$ (see Fig. \ref{fig:stream}).  In the flowing layer (where gradients perpendicular to $\vb{n}$ can be neglected), multiplying \eqref{eq:momentum} by $\vb{n}$  yields
\begin{equation}
    \nabla_{\rm n}\sigma_{\rm n}^{\c,\nu} + \nabla_{\rm n}\sigma_{\rm n}^{\k,\nu} = \beta_{\rm n}^{\nu} + \rho^\nu g_{\rm n},
\label{eq:steady_state}
\end{equation}
where $\sigma_{\rm n}^{\c,\nu}$ and $\sigma_{\rm n}^{\k,\nu}$ denote the contact and kinetic parts of the stress component pointing in the $\vb{n}$, respectively. The gradient $\nabla_{\rm n} = \nabla \cdot \vb{n}$ represents the derivative in $\vb{n}$ direction, and $g_{\rm n}=\vb{g}\cdot\vb{n}$ is the projection of $\vb{g}$ along $\vb{n}$.





In addition to the previously defined partial quantity $\phi^{\nu}$, we define a \emph{kinetic} and \emph{contact stress fraction}, i.e., the stress per unit granular stress, as $f^{\k,\nu} = \sigma_{\rm n}^{\k,\nu}/\sigma_{\rm n}^{\k}$ and $f^{\c,\nu} = \sigma_{\rm n}^{\c,\nu}/\sigma_{\rm n}^{\c}$, as was done in previous contributions \cite{Gray2005theory, Gray2006, Tunuguntla2017}. If the  stress fraction of a constituent exceeds the value $\phi^{\nu}$, we all this condition \emph{over-stress}, else \emph{under-stress}.

The last ingredient of the model is the form taken by $\beta_{\rm n}^{\nu}$; 
similar to Hill and Tan \cite{Hill2014}, we neglect diffusive remixing and propose the following drag terms 
\begin{equation}
\beta_{\rm n} ^{\nu }=\sigma_{\rm n}^{\c,\nu} \nabla_{\rm n} \left( f^{\c,\nu }\right) + \sigma_{\rm n}^{\k,\nu} \nabla_{\rm n} \left( f^{\k,\nu }\right) -\rho ^{\nu }c\left( u_{\rm n}^{\nu }-u_{\rm n}\right)
\label{eq:drag_force}
\end{equation}
The first two terms ensure that, as in Darcy’s law, the percolation process is driven by intrinsic rather than partial stress gradients. The third term is a linear drag law similar to that
provided by Morland \cite{morland1992flow} for the percolation of fluids, where $c$ is an inter-constituent drag coefficient, which we assume to be constant. 
{This form of the drag may appear complicated, but it is both theoretically justified, e.g., \cite{GrayThornton2005,Hill2014} and confirmed in particle simulations \cite{WeinhartLudingThornton2013,BancroftJohnson2021}. Of particular note is the work of \cite{BancroftJohnson2021} as they extended the drag model to give a closed form expression for $c$; however, here we simply give it as to determined parameter, $c$.}

Substituting \eqref{eq:drag_force} in \eqref{eq:steady_state}, the relative percolation velocity is
\begin{equation}\label{eq:stressModel}
    c\phi^{\nu}(u_{\rm n}^{\nu} - u_{\rm n})  = \underbrace{ \dfrac{(f^{\c,\nu} - f^{\k,\nu})}{\rho}\nabla_{\rm n}\sigma^k_{\rm n}}_{\Phi_k} + \underbrace{\dfrac{(f^{\c,\nu} - \phi^{\nu})}{\rho}g_{\rm n}}_{\Phi_g}
\end{equation}
where
$\Phi_k$ and $\Phi_g$ quantify the effects of kinetic and gravity mechanisms. Note that $\Phi_k$ is not present in the original Gray \& Thornton model \cite{Gray2005theory}. Before we compute $\Phi_k$ and $\Phi_g$, it is convenient to obtain the profiles of $(f^{\k,\nu} - \phi^{\nu})$, $(f^{\c,\nu} - \phi^{\nu})$, and $(f^{\c,\nu} - f^{\k,\nu})$ as a function of $\phi^{\nu}$.


\begin{figure}[t]
\includegraphics[scale = 0.8 ]{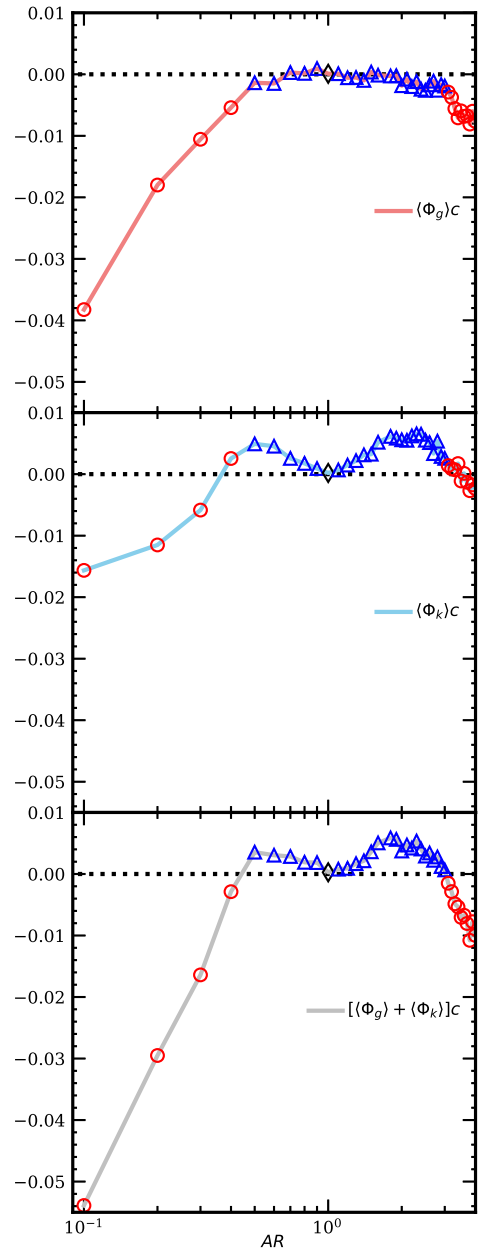}
\caption{{From top to bottom: averaged gravity, $\langle \Phi_g \rangle$ stress; kinetic segregation, $\langle\Phi_k\rangle$; and total stress,$[\langle\Phi_g\rangle + \langle\Phi_k\rangle]$; for formulations see \eqref{eq:stressModel}. The blue triangles are where the spherical particles  segregate inwards and the red spheres outwards. Showing only the combination of the two mechanisms predictions the correct tread.}
} 
\label{fig:forces}
\end{figure}

To perform the analysis related to the quantification of the segregation mechanisms, we averaged the CG-fields on the domain where the density of the mixture is different from zero. Fig.~\ref{fig:prefactors_oblate} 
illustrates the aforementioned profiles for oblate $AR \in \lbrace0.1,\, 0.2,\, 0.4,\, 0.5,\, 0.7\rbrace$ and prolate $AR \in \lbrace1.5,\, 2.3,\, 2.5,\, 3.7,\, 4.0\rbrace$ ellipsoids, respectively. One can see that the values of $(f^{\k,\rm s} - \phi^{\rm s})$ for spheres are positive for all $AR$, indicating that spherical particles support more kinetic stress than their relative concentration. This result agrees with previous observations that spheres possess a better flowability and implies that kinetic sieving alone is unable to explain the change in the segregation direction. 
Moreover, $(f^{\c,\rm s} - \phi^{\rm s})$ is near zero for $AR = 0.5$ and $2.5$ and strictly negative for $AR = 0.1$ and $AR = 4.0$, indicating that spherical particles support equal or less contact stress than their relative concentration. To our knowledge, this is the first time that research evidences this finding. 

{In figure~\ref{fig:prefactors_oblate}, we plot the kinetic $(f^{\mathrm{k},\nu} - \phi^\nu)$ (left),   contact overstresses $(f^{\mathrm{c},\nu} - \phi^\nu)$ (middle), and their difference $(f^{\mathrm{c},\nu} - f^{\mathrm{k},\nu})$ (right). It should be noted that the difference (right) is pre-factor for the kinetic mechanics,  $\Phi_k$ in Eq.~\eqref{eq:stressModel},  and the contact over-stress (right), the gravity mechanism $\Phi_g$ in Eq.~\eqref{eq:stressModel}. Both mechanisms are influenced by the contact {\it over-stress} and that the kinetic and contact overstress are always complementary: mixtures that are contact-{\it overstressed}  are simultaneously kinetic-{\it understressed}, and \textit{vice versa}. Thus, a high contact over-stress 
 also leads to a high prefactor of $\Phi_\mathrm{k}$, that is, the magnitude (not direction) of both mechanisms is highly influenced by the contact stress.} 
 {This is why previously it has been hard to determine if there are one or two different segregation mechanisms.}


We denote the averaging over the flowing layer by $\langle\Phi_g\rangle$ and $\langle\Phi_k\rangle$, respectively.
Fig.~\ref{fig:forces} displays both terms and the averaged total contribution, multiplied by $c$ as a function of $AR$.
Comparing $\langle\Phi_g\rangle$ and $\langle\Phi_k\rangle$ shows that the kinetic mechanism dominates where spheres segregate to the periphery of the drum (blue triangles). However, where the spheres segregate to the core of the mixture (red circles), the gravity mechanism increases as the constituents of the mixture become more different, and the absolute values of $\langle\Phi_g\rangle$ are larger than the $\langle\Phi_k\rangle$ ones. This implies that for positive values of $S$, segregation occurs mainly because of the kinetic stress gradient, and the gravity mechanism is negligible. In contrast, for negative values of $S$, the gravity mechanism drives the segregation, even though $\langle\Phi_k\rangle$ has considerable values. Finally, the averaged total contribution $\langle\Phi_g\rangle + \langle\Phi_k\rangle$ correlates with the behavior of $S$, and the sign changes of the total contribution occur at the same $AR$ values as the sign change in $S$. So, in order to correctly predict the direction of segregation, both mechanisms are required. To the best of the authors' knowledge, this represents the first system where these two effects have been shown to be in competition. 

\section{Conclusions} 

 Our numerical and theoretical analysis accurately explains particle-shape segregation patterns in rotating drums. We present the first evidence that there are two \emph{distinct} segregation mechanisms driven by \emph{over-stress} in the contact and kinetic stresses, as was
suggested by Fan \& Hill \cite{Fan2011}.
In fact, we show that for non-spherical particles, these two mechanisms can act in different directions leading to a competition between the effects of the two. 
This explains the surprising segregation reversal reported by He \textit{et al.} \cite{He2019}. Particularly, in rotating drums, the segregation intensity varies non-monotonically as a function of $AR$, and at specific points, the segregation direction changes for both prolate and oblate spheroids. Remarkably, our analysis predicts the location of these transition points quantitatively, quantifying the relative momentum interchange between the species.
Consistent with previous results \cite{Fan2011, Tunuguntla2017} we found that the kinetic mechanism is dominant for (almost) spherical particles.  For moderate $AR$ values, the kinetic mechanism is responsible for spherical particles segregating to the periphery of the drum, and the gravity mechanism plays only a minor role. Whereas, at the extreme values of $AR$, the gravity mechanism increases rapidly and dominates its kinetic counterpart.

D.H.\ acknowledges Asociaci\'on de Amigos de la Universidad de Navarra, MercuryLab, and Fundación Caja Navarra-Caixa. This work was partially funded by Ministerio de Economía y Competitividad (Spanish Government) through the Project PID2020-114839GB-I00 MINECO/AEI/FEDER, UE. R.C.H.\ acknowledgments the European Union's Horizon 2020 Research and Innovation programme under the Marie Sklodowska-Curie grant agreement CALIPER No 812638.  A.T.\ acknowledges the support of the Dutch Research Council, STW-Vidi project 13472. T.W.\ acknowledges the support of the Dutch Research Council, NWO-TTW grants 15050 and 16604. Finally, we acknowledge the contributions of I.F.C. Dennissen, who started the superquadrics implementation and the early simulations on this topic. 
\cite{Denissen2019}.

\bibliography{MyBibText}

\end{document}